\documentstyle[preprint,aps,floats,epsfig]{revtex}

\newcommand{\be}{\begin{equation}}
\newcommand{\ee}{\end{equation}}
\newcommand{\bea}{\begin{eqnarray}}
\newcommand{\eea}{\end{eqnarray}}
\newcommand{\bml}{\begin{mathletters}}
\newcommand{\eml}{\end{mathletters}}

\begin{document}
\preprint{DTP/00/31, hep-th/0004101}
\draft
\tighten

\title{Black string instabilities in anti-de Sitter space}
\author{ Ruth Gregory\footnote{E-mail address:
        \texttt{R.A.W.Gregory@durham.ac.uk}}}
\address{Centre for Particle Theory, 
         Durham University, South Road, Durham, DH1 3LE, U.K.}
\date{\today}
\setlength{\footnotesep}{0.5\footnotesep}
\maketitle

\begin{abstract}
We show how to extend the usual black string instability of vacuum or
charged black p-branes to the anti-de Sitter background. The
string fragments in an analogous fashion to the $\Lambda=0$ case,
the main difference being that instead of a periodic array of black holes
forming, an accumulation of ``mini'' black holes occurs towards the
AdS horizon. In the case where the AdS space is of finite extent, 
such as an orbifold compactification, we show how the instability 
switches off below a certain compactification scale.

\
\end{abstract}

\pacs{PACS numbers: 04.50.+h, 11.25.Mj, 11.25.Db \hfill hep-th/0004101}


It has been known for some time that the extended event horizon of
a black string or $p$-brane in higher-dimensional gravity is unstable to 
fragmentation into black holes \cite{GL1}, and that this instability 
extends to a much broader range of charged black holes in string theory -- 
the only exception being extremal solutions \cite{GL2}. There are
several ways of understanding this instability intuitively: the event
horizon of a black hole can be thought of as analogous to a soap bubble --
it has a `tension' ($\kappa$, the surface gravity), and one
therefore expects a cylindrical horizon to cease being a stable
solution once its length reaches some critical value. Another
(related) quantum description is that we can associate an entropy to an event
horizon, proportional to its area, and a simple calculation shows
that for a cylindrical horizon, there exists a length above which it
becomes entropically favourable for the mass to localise in a spherical
black hole. These simple arguments can be backed up by a full perturbative
analysis of the spacetime of the extended solution, where it is straightforward
(if calculationally involved) to show that there exists a transverse-tracefree
(TTF) metric perturbation, which is a pure tensor mode in the spacetime
orthogonal to the string or brane \cite{GL1,GL2}.

These instability arguments were made purely within the context of
asymptotically locally flat spacetimes -- in other words, in the
absence of a cosmological constant. Recent interest on the other hand
has focussed on
backgrounds which are not vacuum, but rather negative false vacuum or 
anti-de Sitter (AdS) spacetimes. Such backgrounds arise in the near-horizon
limit of certain D-branes, and the general results of \cite{GL1,GL2} have
been used for example
in the context of localisation of AdS$_5$ black holes on the $S^5$
component of the near horizon geometry of the D3-brane \cite{BDHM}. 

A particularly exciting application of AdS spacetime to the `real world'
has arisen more recently in the context of superstring cosmological scenarios
resulting from the Horava-Witten \cite{HW} compactification of M-theory.
In this case, spacetime is effectively five-dimensional below the
Grand Unified scale, \cite{LOW}, and our universe arises as a domain wall, 
or brane, at its boundary. The idea of our universe as a defect in higher 
dimensions is not new (see e.g.\ \cite{UDW}), however, 
Randall and Sundrum \cite{RS} emphasized
out the exciting phenomenological implications of this set-up --
not only is gravity essentially four-dimensional at reasonably large
length scales, but there is also the possibility of observing fundamental 
spin-2 particles at LHC (originally pointed out in \cite{AH}).
It is this former feature which is relevant to the current discussion. 

If our universe is to be regarded as a domain wall
in five-dimensional AdS spacetime, a natural question is what is the
gravitational field of a test particle in our universe? In linearized
theory, the metric on the wall was shown to be exactly what one would
obtain from four-dimensional Einstein gravity, modulo short range corrections
due to the massive Kaluza-Klein (KK) graviton modes \cite{GT}. 
However, it is obviously
of interest to determine the full non-perturbative five-dimensional `vacuum'
solution corresponding to a black hole in the brane. This question was
explored in \cite{CHR} (see also \cite{CCEH}), where the 
instability of the Einstein black string 
was used to argue a `cigar' geometry of the event horizon of such a black
hole. 

Five dimensional AdS spacetime can be written in horospherical
coordinates as:
\be
ds^2 = e^{-2kz} \left [ \eta_{\mu\nu} dx^\mu dx^\nu \right ] - dz^2
\label{adsmet}
\ee
where the cosmological constant is $-6k^2$. The Randall-Sundrum
spacetime simply places a wall of tension $6k$ at $z=0$, and a second
wall of tension $-6k$, if required, at some $z = z_c$. This has the
effect of making the metric reflection symmetric around $z=0$ and $z=z_c$.
In either case, it is straightforward to construct the black string, since
the flat metric $\eta_{\mu\nu}$ can be replaced by any Ricci-flat
metric $g_{\mu\nu}$, for example the four-dimensional Schwarzschild
solution. The properties of this Randall Sundrum (RS) black string 
were explored by Chamblin et.\ al.\ 
in \cite{CHR}. In particular, it was argued that the 
Einstein black string instability would come into play near the AdS
horizon indicating a cigar type of event horizon.
On the other hand, an alternative approach put forward by
Emparan et.\ al.\ \cite{EHM}, in which the C-metric is used to 
construct an exact black hole on a 2+1 dimensional universe at the
boundary of AdS$_4$, appears to indicate that the horizon of the
black hole will appear more as a pancake localised on the wall,
indeed these authors used the Einstein black string instability to
argue this time the pancake geometry in higher dimensions!
Clearly, since the arguments used in these papers involved an application
of the instability in a situation other than that in which it was derived,
a correct calculation would be useful in such a discussion.

Let us start by recalling the instability for the black string
in Einstein gravity. The metric of the string is given by simply
adding an extra flat direction to the Schwarzschild metric:
\be
ds^2 = \left ( 1 - {2GM\over r} \right ) dt^2 -
\left ( 1 - {2GM\over r} \right )^{-1} dr^2 - r^2 d\Omega_{I\!I}^2 -dz^2
\label{bstr}
\ee
Writing the perturbation of the metric in the usual fashion as
$g_{ab} \to g_{ab}+h_{ab}$, and choosing
the TTF gauge for $h_{ab}$ (i.e.\ $h^a_a=0=
h^a_{b;a}$), then the perturbation $h_{ab}$ satisfies the Lichnerowicz
equation:
\be
\Delta_L h_{ab} = \left [ \delta^c_a \delta^d_b \Box + 2R^{\;c\;d}_{a\;b}
\right] h_{cd} = 0
\ee
It was shown in \cite{GL1} that this could be interpreted 
in a Kaluza-Klein spirit as a four-dimensional problem, with the
perturbation splitting into scalar, vector and tensor modes. The
scalar and vector modes were shown to be zero for any unstable mode,
and so the problem reduced to a four-dimensional tensor perturbation
\be\label{4deom}
\left ( \Delta_L^{(4)} + m^2 \right ) h_{\mu\nu} = 0
\ee
where the $z$-dependence of the perturbation $e^{imz}$ simply introduces
a mass term for the tensor mode $h_{\mu\nu}$. Since all pure gauge
perturbations satisfy the massless equation, any solution to the above
massive Lichnerowicz equation must therefore be physical and correspond
to an instability of the spacetime. 

A single s-wave mode was found in \cite{GL1} which had the form
\be
h_{\mu\nu} = e^{\Omega t} \left [ \matrix{ h_0 & h_1 & 0 & 0 \cr
h_1 & h_2 & 0 & 0\cr 0 & 0 & K & 0 \cr 0 & 0 & 0 & K\sin^2\theta \cr}
\right ]
\ee
Where $h_0,h_1,h_2$ and $K$ are all related via the TTF gauge conditions,
and $h_{\mu\nu}$ has the interpretation of a longitudinal graviton mode
which perturbs the event horizon, making it ripple with a characteristic
wavelength leading to the interpretation of the instability as being
a process which leads ultimately to the fragmentation of the cylindrical
event horizon into a line, or array, of black holes. 

To sum up, the key (mathematical) features of this instability are 
that {\it i)} it has the form of a tensor perturbation from the
four-dimensional point of view; {\it ii)} this tensor mode satisfies
the TTF gauge conditions\footnote{In the context of the Randall-Sundrum
braneworld scenario, this gauge, which is simultaneously 
GN (Gaussian-Normal : $g_{zz}=1$, $g_{z\mu}=0$)
{\it and} TTF, is often called the RS gauge.}; and {\it iii)} it
satisfies a {\it massive} four-dimensional Lichnerowicz equation.

Now let us consider the black string in AdS spacetime, which has the metric
\be
ds^2 = a^2(z) \left [ \left ( 1 - {2GM\over r} \right ) dt^2 -
\left ( 1 - {2GM\over r} \right )^{-1} dr^2 - r^2 d\Omega_{I\!I}^2 
\right ] -dz^2 
\label{ads}
\ee
The function $a(z)$ has been left general in this expression to allow
for the simultaneous consideration of instabilities of pure AdS black strings,
$a(z) = e^{\pm kz}$, or the RS black string, $a(z) = e^{-k|z|}$ (which
may or may not have an additional negative tension domain wall at some other
value of $z$). Note that although this looks from the four-dimensional
point of view like a Schwarzschild black hole of mass $M$, the $r$-coordinate
is really a `comoving' coordinate, and the `ADM' mass as measured by
an observer in a particular $z_0$ plane is in fact $M_0 = M e^{-kz_0}$.
The spacetime becomes singular as the AdS horizon is approached
\cite{CHR}.

Clearly, the only 
difference between the metrics (\ref{ads}) and (\ref{bstr}) is the conformal
or warp factor in front of the four-dimensional Schwarzschild part. This
is crucial however, in that this warp factor represents the
introduction of the negative cosmological constant, and is responsible
for the {\it de facto} compactification of the spacetime (even though the extra
dimension has infinite extent)  in the single domain wall universe as
explored by Squires and Visser \cite{EXO}. Unlike the standard
Kaluza-Klein compactification of spacetime, in which the effective
theory is that of coupled massless spin two, one and zero particles, together 
with a tower of massive spin two states, the `exotic'
compactification induced by the AdS spacetime, appears to
leave only a massless spin two excitation, the four-dimensional Einstein
graviton, and a continuum of massive KK states. We might therefore expect
that the instability argument would be severely modified, and calculationally
a great deal more involved, nevertheless, for the moment
let us proceed innocently\footnote{If we wish to be less innocent, we
can write the perturbation equations in a GN but not TTF gauge, and 
demonstrate directly that in fact the RS gauge suffices on-shell, since in
the absence of additional matter perturbations, the solutions for the
trace and divergence of well-behaved perturbations must vanish.}, 
and regard this warp factor 
as simply representing the exotic compactification of spacetime,
and look for an tensor instability of a similar form to
the vacuum one.

Writing $R^{(4)}_{\mu\nu\lambda\rho}$ for the Riemann tensor
of the four-dimensional Schwarzschild metric, and taking a
RS (four-dimensional TTF tensor) perturbation, we find that the perturbation
equations for the metric (\ref{ads}) reduce to 
\be
a^{-2} \left ( \Box^{(4)} h_{\mu\nu} + 2 R^{(4)}_{\mu\lambda\nu\rho} 
h^{\lambda\rho} \right ) - h''_{\mu\nu} + 2 \left ( {a''\over a} 
+ \left({a'\over a}\right)^2 \right ) h_{\mu\nu} = 0
\ee
where the greek spacetime indices have been raised by the four-dimensional
Schwarschild metric {\it without} the warp factor, and the wave operator
similarly represents the four-dimensional Schwarzschild wave operator.

Now, setting $h_{\mu\nu} = \chi_{\mu\nu} u_m(z)$, where 
\be
u_m(z) = {\cal A}\, {\rm J}_2 \left ( {m\over k} e^{kz} \right ) 
- {\cal B}\, {\rm N}_2 \left ( {m\over k} e^{kz} \right ) 
\label{umdef}
\ee
with the coefficients ${\cal A}$ and ${\cal B}$ being chosen so that
the perturbation satisfies the required boundary conditions for either
a pure AdS black string (${\cal B}=0$), or a RS black string
(${\cal A}{\rm J}_1\left ( {m\over k} \right ) = {\cal B}
{\rm N}_1 \left ( {m\over k} \right )$). Then in either case
\be
- h''_{\mu\nu} + 2 \left ( {a''\over a} + \left({a'\over a}\right)^2 \right )
h_{\mu\nu} = a^{-2} m^2 h_{\mu\nu}
\ee
and hence $\chi_{\mu\nu}$ satisfies the equation of motion
\be
\left ( \Delta_L^{(4)} + m^2 \right ) \chi_{\mu\nu} = 0
\ee
where the $z$-dependence of $u_m(z)$, has simply introduced a mass term for
$\chi_{\mu\nu}$, in direct analogy to the flat extra dimension case
(\ref{4deom}).
\begin{figure}
\centerline{\epsfig{file=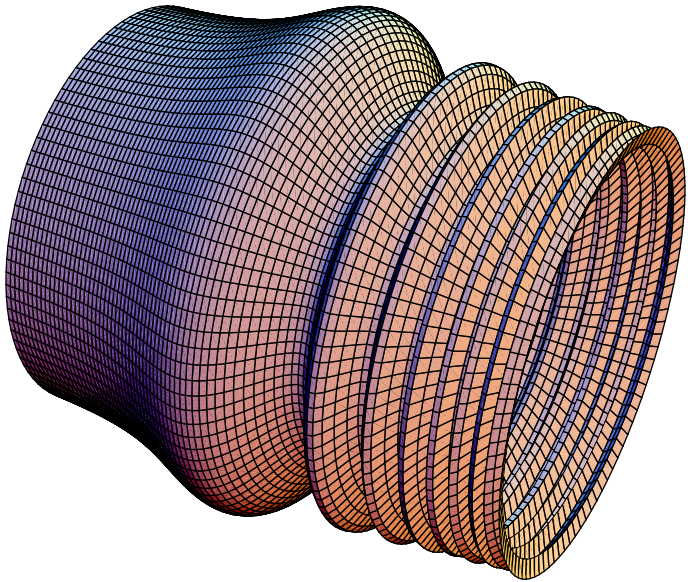,width=11cm}}
\caption{The event horizon of the perturbed AdS black string in horospherical
coordinates.}
\label{fig:adsins}
\end{figure}
Therefore we can simply read off the solution for $\chi_{\mu\nu}$
from \cite{GL1}, and see the form of the instability. 
For $GM=1$, the instability
exists in the range $0<m< 0.45$, with the most favoured instability
(i.e.\ the one with the shortest half-life) having $m\simeq0.2$.
For general $M$, the perturbation mass varies as $M^{-1}$.

\begin{figure}
\centerline{\epsfig{file=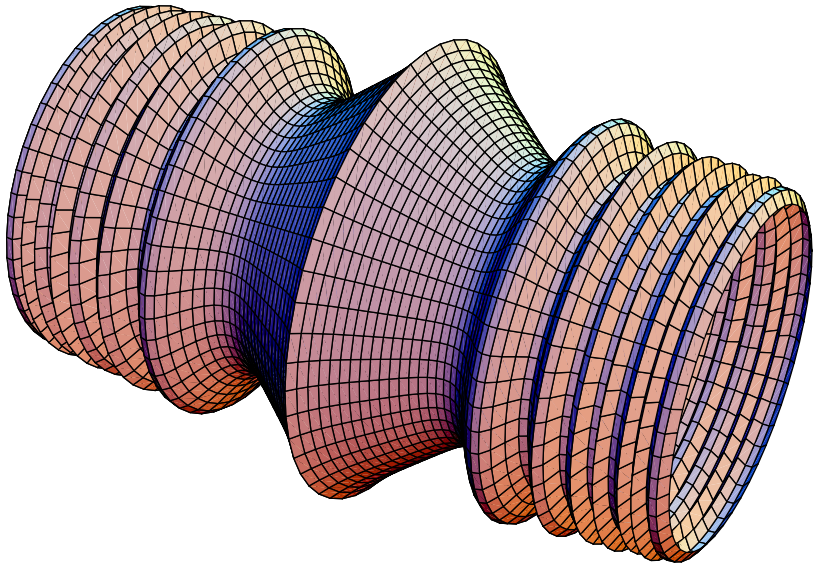,width=12.8cm}}
\caption{The event horizon of the perturbed RS black string.}
\label{fig:hwall}
\end{figure}

An image of the effect of the instability on the horizon of the pure AdS string 
is shown in figure \ref{fig:adsins}, and for the Randall-Sundrum string
in figure \ref{fig:hwall}. Note that in this representation of the instability,
we are in `comoving' coordinates, i.e.\ the unperturbed black string looks 
cylindrical. 
Notice how in each case
the instability accumulates towards the AdS horizon $z\to\infty$.
From the asymptotics of the Bessel functions, we see that successive
zeros of $h_{\mu\nu}$ are separated by $\pi e^{-kz}/m$, and hence the `mass'
of black string contained in these seedling black holes decreases roughly
as ${\cal M}_s \sim \pi k M e^{-2kz} /m$. Thus we see that the instability,
if it proceeds into fragmentation of the black string, will generate an 
accumulation of mini-black holes towards the AdS horizon. Although such a
conclusion seems at first alarming, in that Planck scale black holes will
rapidly dominate, it is only fair to point out that since the original black 
string solution is singular at the AdS horizon \cite{CHR},
perhaps this pathological behaviour of the instability is simply
mirroring this fact. In any case, we can always introduce a regulator
brane at some finite $z_c$ (see below) to act as a cut-off for such singular 
behaviour.

Clearly the instability differs for lower values of $z$, where the
wall of the Randall-Sundrum string makes its presence felt. For the
pure AdS string, the instability is suppressed in the region $z<0$, 
or near the boundary of AdS spacetime. This corresponds to the
region in which the curvature of the AdS spacetime is large 
compared to the Schwarzschild curvature. For the RS spacetime, 
the first node of the instability occurs at a proper distance 
$z_n$ of approximately 2 from the wall for $GM=k=1$ and $m=m_{\rm max}=0.45$,
indicating the typical scale of the central black hole after fragmentation. 
Since the horizon of the black holes in these units is 2, this seems to 
indicate neither a pancake geometry nor a cigar one for black strings 
with masses on a similar scale to the AdS curvature (as might be expected). We 
can however use the asymptotics of the Bessel functions to
explore the possible geometry of black holes on the brane in the
asymptotic r\'egimes of very large and very small masses. 
For $m/k \ll 1$, this first node of the instability occurs for
$z \simeq -{1\over k}\ln ({m\over k})$, i.e.\ $z_n \propto (\ln M)/k$
for $M\gg 1$, which indicates a `stottie'\footnote{A regional bread, 
of a similar shape to the pancake or tortilla, but somewhat thicker.} 
shape for the event horizon.
On the other hand, for $m/k \gg1$ we obtain $z \propto \pi/m$,
or $z\propto M$. This indicates that very small mass black holes are
roughly hyperspherical, which is approximately what we might expect
since the black hole is at a scale where the AdS curvature is
negligible, and we might expect it to look like a five-dimensional
black hole. These results  are in accord with the exact 2+1-dimensional
calculations of \cite{EHM}, as well as with an analysis of the propagator 
off the brane in \cite{GKR}.

It is interesting to explore what happens if we have a finite
fifth dimension, as in the original Randall-Sundrum scenario.
In this case, the presence
of a second wall at $z_c$ introduces an additional restriction on
the eigenfunctions $u_m(z)$ :
\be
h_m'(z_c) + 2k u_m(z_c) = 0 \;\;\; \Rightarrow \;\;\;
{\rm J}_1 \left( {m\over k}e^{kz_c} \right) {\rm N}_1 \left ({m\over k}\right )
= {\rm J}_1 \left({m\over k}\right){\rm N}_1 \left ({m\over k}e^{kz_c} \right)
\ee
Clearly, if $kz_c$ is not sufficiently large, there will be no allowed
value of $m$ for which the eigenfunction $u_m$ exists. For small $m$,
$kz \propto \ln ({k\over m})$, and hence for large mass black holes,
$e^{kz_c} \geq 2kGM$ for the existence of the instability.

We can also straightforwardly derive the form of the instability for
the quasi-localized gravity model of \cite{GRS}, where the central
positive tension wall is flanked by two negative tension walls at $z=z_c$ in
such a fashion that the spacetime is Minkowskian outside the system.
The eigenfunctions have the form of (\ref{umdef}) inbetween the walls, 
and the conventional oscillatory form in the exterior
flat spacetime. In this case, for $M \ll e^{kz_c}/k$ the instability
proceeds primarily as for the RS string, giving a picture much like
figure \ref{fig:hwall}, but for $M \gg e^{kz_c}/k$, the first node of the
instability lies outside the wall system, and we have a central,
slightly squashed, black hole forming on the walls, with the usual
line of five-dimensional black holes extending off to infinity.

Finally, we note that although we have focussed, mainly for the
purposes of clarity, on the case of a black string in AdS$_5$
(or a slice thereof), it is clear how to generalize this to black
strings, or indeed branes, in higher dimensional AdS spacetimes.
The key feature required is a block diagonal form of the metric, 
which may or may not depend nontrivially on the extra dimensions.
For example, a six-dimensional model in which the universe is a `vortex'
\cite{G} (see also \cite{CK}), rather than a wall, can be used 
to analyse the instability of a black membrane. In this case, the 
analogues of the $u_m$'s have been investigated in \cite{GS}.  Similarly,
for higher dimensional branes, the smooth solutions of Olasagasti
and Vilenkin \cite{OV} can be used to model the behaviour of the massive
graviton modes.

\section*{Acknowledgements}

I would like to thank Peter Bowcock, Roberto Emparan,
Paul Mansfield and Simon Ross
for useful conversations and comments on the manuscript.
This work was supported by the Royal Society.

\def\book#1[[#2]]{{\it#1\/} (#2).}
\def\apj#1 #2 #3.{{\it Astrophys.\ J.\ \bf#1} #2 (#3).}
\def\atmp#1 #2 #3.{{\it Adv.\ Theor.\ Math.\ Phys.\ \bf#1} #2 (#3).}
\def\cmp#1 #2 #3.{{\it Commun.\ Math.\ Phys.\ \bf#1} #2 (#3).}
\def\comnpp#1 #2 #3.{{\it Comm.\ Nucl.\ Part.\ Phys.\  \bf#1} #2 (#3).}
\def\cqg#1 #2 #3.{{\it Class.\ Quant.\ Grav.\ \bf#1} #2 (#3).}
\def\grg#1 #2 #3.{{\it Gen.\ Rel.\ Grav.\ \bf#1} #2 (#3).}
\def\jmp#1 #2 #3.{{\it J.\ Math.\ Phys.\ \bf#1} #2 (#3).}
\def\ijmpd#1 #2 #3.{{\it Int.\ J.\ Mod.\ Phys.\ \bf D#1} #2 (#3).}
\def\mpla#1 #2 #3.{{\it Mod.\ Phys.\ Lett.\ \rm A\bf#1} #2 (#3).}
\def\ncim#1 #2 #3.{{\it Nuovo Cim.\ \bf#1\/} #2 (#3).}
\def\npb#1 #2 #3.{{\it Nucl.\ Phys.\ \rm B\bf#1} #2 (#3).}
\def\phrep#1 #2 #3.{{\it Phys.\ Rep.\ \bf#1\/} #2 (#3).}
\def\pla#1 #2 #3.{{\it Phys.\ Lett.\ \bf#1\/}A #2 (#3).}
\def\plb#1 #2 #3.{{\it Phys.\ Lett.\ \bf#1\/}B #2 (#3).}
\def\pr#1 #2 #3.{{\it Phys.\ Rev.\ \bf#1} #2 (#3).}
\def\prd#1 #2 #3.{{\it Phys.\ Rev.\ \rm D\bf#1} #2 (#3).}
\def\prl#1 #2 #3.{{\it Phys.\ Rev.\ Lett.\ \bf#1} #2 (#3).}
\def\prs#1 #2 #3.{{\it Proc.\ Roy.\ Soc.\ Lond.\ A.\ \bf#1} #2 (#3).}

\end{document}